\documentclass[useAMS]{mn2e}

\usepackage{graphicx}	
\usepackage{amsmath}	
\usepackage{amssymb}	
\usepackage{multicol}       
\usepackage{bm}		
\usepackage{pdflscape}	
\usepackage[utf8]{inputenc}
\usepackage[T1]{fontenc}
\usepackage{ae,aecompl}

\title[Statistical analysis of {\it GAIA} wide binaries] {Statistical analysis of the gravitational anomaly in {\it Gaia}
  wide binaries.} 

\author[X. Hernandez, V. Verteletskyi, L. Nasser and A. Aguayo-Ortiz] {X. Hernandez$^{1}$, V. Verteletskyi$^{2}$,
  L. Nasser$^{3}$ and A. Aguayo-Ortiz$^{1}$\\ 
$^{1}$ Instituto de Astronom\'{\i}a, Universidad Nacional Aut\'{o}noma de M\'{e}xico,
Apartado Postal 70--264 C.P. 04510 M\'exico D.F. M\'exico. \\
$^{2}$ IMC Trading, Amstelveenseweg 500, 1081 KL Amsterdam, The Netherlands.\\
$^{3}$ Department of Science and Mathematics, Columbia College, Chicago, IL 60605, USA. \\
}

\date{Released 20/09/2023}

\begin{document}

\label{firstpage}

\maketitle

\begin{abstract}
  The exploration of the low acceleration $a<a_{0}$ regime, where $a_{0}=1.2 \times 10^{-10}$m s$^{-2}$ is the acceleration scale of MOND around which
  gravitational anomalies at galactic scale appear, has recently been extended to the much smaller mass and length scales of local wide binaries 
  thanks to the availability of the {\it Gaia} catalogue. Statistical methods to test the underlying structure of gravity using large samples of
  such binary stars and dealing with the necessary presence of kinematic contaminants in such samples have also been presented. However, an alternative
  approach using binary samples carefully selected to avoid any such contaminants, and consequently much smaller samples, has been lacking a formal
  statistical development. In the interest of having independent high quality checks on the results of wide binary gravity tests, we here develop a
  formal statistical framework for treating small, clean, wide binary samples in the context of testing modifications to gravity of the form
  $G \to \gamma G$. The method is validated through extensive tests with synthetic data samples, and applied to recent {\it Gaia} DR3 binary star
  observational samples of relative velocities and internal separations on the plane of the sky, $v_{2D}$ and $r_{2D}$, respectively. Our final results
  for a high acceleration $r_{2D}<0.01$pc region are of $\gamma=1.000 \pm 0.096$, in full accordance with Newtonian expectations. For a low
  acceleration $r_{2D}>0.01$pc region however, we obtain $\gamma=1.5 \pm 0.2$, inconsistent with the Newtonian value of $\gamma=1$ at a $2.6 \sigma$
  level, and much more indicative of MOND AQUAL predictions of close to $\gamma=1.4$.
\end{abstract}


\section{Introduction}

In the context of the debate surrounding the identification of low acceleration gravitational astronomical anomalies as either the
result of a change in gravity at those scales, or as indication of the existence of a dominant dark matter component, wide binaries have
been identified as capable of providing relevant independent insights, Hernandez et al. (2012). {  Solar mass star binaries on circular orbits
with separations larger than 0.035 pc (7000 au) lie in the regime where accelerations fall below $ a_{0}$, where
$a_{0}=1.2 \times 10^{-10}$m s$^{-2}$ is the characteristic acceleration scale of MOND, an indicative threshold at which observed galactic
dynamics show the above mentioned gravitational anomalies, e.g. Milgrom (1983), Lelli (2017).

Under a modified gravity interpretation one expects the appearance of gravitational anomalies at accelerations larger than $a_{0}$ by a factor
of a few, due to the presence of a smooth transition between regimes. In the particular case of the wide binaries treated, gravitational
anomalies would be expected at separations smaller than the 0.035 pc mentioned above for an additional reason: the mean total masses
per binary system are of only 1.5 $M_{\odot}$. Indeed, recently Hernandez et al. (2022), Chae (2023a) and Hernandez (2023) using {\it Gaia}
wide binaries have reported gravitational anomalies appearing at separations above 0.01 pc (2000 au). }

Given the inferred local volume density of dark matter, its total content expected within a wide binary orbit is negligible in
comparison to the masses of the stars themselves. Further, given the assumed velocity dispersion of the hypothetical dark matter
particles of the Milky Way halo of 160 km s$^{-1}$, clustering on scales with dynamical equilibrium velocities of $<1$km s$^{-1}$,
as applies to local wide binaries, would require orders of magnitude of cooling, for a component which by construction must be
dissipationless. Thus, any gravitational anomaly of the type encountered at galactic scales and beyond, found in wide binary stars,
cannot comfortably be ascribed to the presence of dark matter.

There are details to be taken into account when performing such a wide binary gravity test. Crucially, because the orbital timescales
of the systems in question are of many thousands of years, the test can only be undertaken statistically by examining large samples
of wide binaries and comparing observed distributions of relative internal velocities to various competing models, e.g.
Hernandez et al. (2012), {  Pittordis \& Sutherland (2018)}, Banik \& Zhao (2018), Hernandez et al. (2019), Acedo (2020), Pittordis
\& Sutherland (2023), Hernandez (2023) and Chae (2023a). Another key concern is the presence of kinematic contaminants in
local wide binary samples, cases where the two stars of a candidate binary do not in fact form a bound pair, but only a close
flyby event e.g. Pittordis \& Sutherland (2019), and hidden tertiaries, cases where one or both of the stars in a bound binary might
in fact be an unresolved binary itself, e.g. Banik \& Zhao (2018), Clarke (2020). In either of the above cases, the observed relative
velocity between both identified components will be the result of the internal gravitational attraction between both stars, and also
of unrelated physical ingredients; the initial conditions of the hyperbolic flyby or the internal dynamics of the unresolved binary.

One approach has been to attempt to model one (e.g. hidden tertiaries in Chae 2023a) or both of these kinematic contaminants
(e.g. Pittordis \& Sutherland 2023) and account for their effects so as to identify the underlying behaviour of gravity after having
modelled out the contribution of kinematic contaminants. These studies have in one case recently reported the presence of a
gravitational anomaly consistent with MOND appearing for separations on the plane of the sky, $r_{2D}$, larger than $r_{2D}>0.01$ pc,
while calibrating a hidden tertiary model in the Newtonian high acceleration $r_{2D}<0.01$ pc regime and carefully excluding from the sample
flyby events using isolation and relative radial velocity cuts, Chae (2023a). In the other case, attempting to model simultaneously
both sources of kinematic contaminants, and not considering the high acceleration Newtonian region for calibration or consistency checks,
Pittordis \& Sutherland (2023) report a better fit to Newtonian gravity than to {  a modified gravity model tested}, looking
only at the $r_{2D}>0.01$pc regime.

An independent approach is to attempt a thorough cleaning of all kinematic contaminants before performing any gravity test with a
wide binary sample, where very careful selection strategies are required, leading to much smaller samples than the ones mentioned
above. This has been reported in Hernandez et al. (2022) and Hernandez (2023), showing the appearance of a gravitational anomaly
at the same $r_{2D}$ threshold as reported by Chae (2023a), although lacking any formal statistical analysis of the details of such
an anomaly.

{  The stringent requirements of a clean sample limit strongly the final numbers of binaries considered, yielding almost two orders
of magnitude fewer binaries than those used in large samples where removal of kinematic contaminants is much less thorough, e.g.
Pittordis \& Sutherland (2023) or Chae (2023). However, at the expense of numbers, there is a significant increase in gain in certainty
that the binaries included do in fact represent the physics one is trying to asses, e.g. Hernandez (2023), Chae (2023b). This makes both
approaches complementary, largely independent and valuable avenues towards a final answer on this subject.}

The present paper develops and presents an application of a formal statistical method to infer a gravity model {  where Newton's
constant is re-scaled by a fixed factor, as expected for Solar Neighbourhood wide binaries under MONDian models (e.g. Banik \& Zhao 2018)},
parameterised as $G \to \gamma G$, using a small clean sample of local wide binaries from the {\it Gaia} DR3. Optimal values of $\gamma$ relevant to
both the high acceleration $r_{2D}<0.01$pc and the low acceleration $r_{2D}>0.01$pc regimes are obtained. This analysis includes
a full probabilistic treatment of the probability density functions (PDFs) for the two projection angles of the wide binary orbits
involved, the sampling of a distribution of semi-major axes, ellipticities, orbital phase angles and relative velocity
errors.

Section (2) summarises the sample selection strategy and first order results, Section (3) develops the probabilistic treatment
of the problem, which is applied in Section (4) to the {\it Gaia} DR3 sample previously described. Lastly, Section (5) includes
a final discussion of the results obtained and their implications.

\begin{figure}
 \vskip 0pt
 \hskip -5pt
 \includegraphics[height=7.0cm,width=8.5cm]{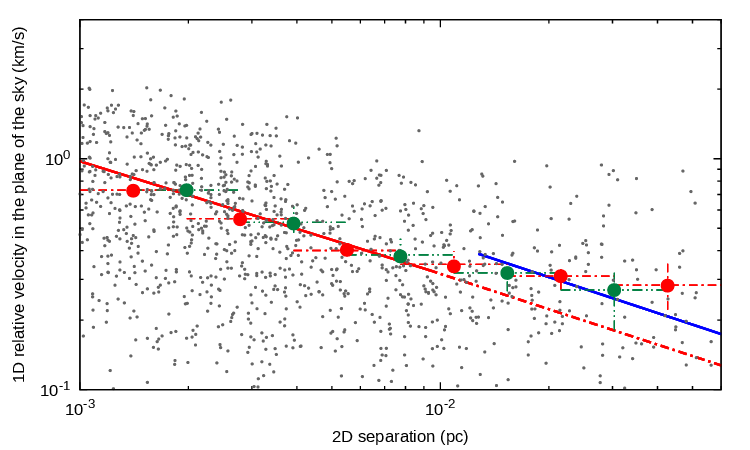}
 \caption{Log-Log plot of 1D relative velocities between the two components of each wide binary in the sample on the plane of the sky as a function of
   the observed separation of the two components on the same plane, small grey dots, RA and Dec. measurements appear separately. The
   large dots with error bars give binned averages for the 1D relative velocities, green and red for RA and Dec., respectively.
   Horizontal bars show the extent of the bins and vertical ones the $1\sigma$ confidence intervals on the quantities given. The red
   line shows a $v_{2D} \propto r_{2D}^{-1/2}$ fit, which is followed by the date in the $r_{2D}<0.01$pc range. In the low acceleration
   $r_{2D}>0.01$ pc range, a similar fit is shown which appears a factor of 1.62 above the previous one. If this factor is interpreted
   as due to an effective change in $G \to \gamma G$, one would infer $\gamma=1.27$ from this rough comparison.}
 \end{figure}

\section{Sample selection and preliminary results}

The DR3 {\it Gaia} wide binary sample analysed in this paper is a small extension of the sample described and treated in
Hernandez (2023), where all the details describing this sample can be found. In summary, the sample comprises 667
wide binaries within a distance of $D <125$ pc from earth and a minimum signal-to-noise ratio in parallax of $(S/N)_{\varpi}=100$,
where an initial binary candidate selection criteria, following  El-Badry \& Rix (2018), selects a list of main sequence stellar
pairs such that twice the separation on the plane of the sky, $2r_{2D}$, is less than the separation along the line of sight, to within
three times the $1\sigma$ confidence interval of this last quantity. Such binary companion candidates are sought up to a separation on the
plane of the sky $r_{2D}<0.5$ pc. This initial list returns many binary candidates with shared stars, that are removed to construct a
catalogue where each binary system is isolated from all other {\it Gaia} sources to within $0.5$ pc, almost an order of magnitude larger
than the largest internal separation used of $r_{2D}=0.06$ pc. Then, quality cuts are imposed to leave only binaries where both stars have
$R_{p}$, $G$ and $B_{p}$ signal-to-noise values $>20$, and where both stars have a reported radial velocity measurement in the catalogue.

Requiring a radial velocity measurement for all stars allows to calculate all astrometric corrections including not only full spherical
geometry corrections, but also perspective effects, e.g. Smart (1968), and also ensures each individual star has a high quality single
star spectroscopic, photometric and astrometric {\it Gaia} solution, something which strongly eliminates hidden tertiaries. Indeed,
many sources lack a reported radial velocity measurement precisely because of a poor single stellar solution.

Next, a series of cuts are introduced to further reduce to a minimum the probability of any kinematic contamination remaining
in the sample. Following Belokurov et al. (2020) and Penoyre et al. (2020) a careful selection of a region of the main sequence in
the CMD diagram below the old turn-off points of the stars obtained is performed (see Hernandez 2023), for all stars involved. This
excludes photometric binaries, and minimises the probability of keeping unresolved hidden binaries. Indeed, the two authors above estimate
through extensive simulations reproducing {\it Gaia} DR2 observational constrains, that the probability of keeping unresolved hidden
tertiaries in samples out to 1kpc is below 5\%, after selecting the CMD region described above and imposing a {\it Gaia} 
{\it RUWE} single star solution quality index cut of $<1.4$. Here we impose a much more stringent {\it RUWE}$<1.2$ limit, and
remain within a much smaller distance of only $D<125$pc, using the more accurate DR3.

Finally, a cut in the upper allowed value of the {  CLASSPROB\_DSC\_COMBMOD\_BINARYSTAR} {\it Gaia} DR3 parameter, henceforth
$B_{P}$, of $B_{P}<0.4$ is introduced. This parameter gives an assessment of the likelihood that a single {\it Gaia} source might
be in fact a binary star, not an actual statistical probability, but at present only a qualitative assessment (G. Gilmore, private
communication). For this reason this last cut was relaxed from the $B_{P}<0.2$ used in Hernandez (2023), which allows for an increase
of about $50\%$ on the total final numbers of binary systems included. Still, the use of all the above parameters sequentially
ensures that final average data quality values are well above the individual thresholds introduced. The relevant observational parameters of
the sample used are given in Table (1), where we can see for example, final mean signal-to-noise in parallax of close to 900,
mean values of {\it RUWE} of 1.01 and mean values of $B_{P}$ of 0.12 for the samples used. All cuts on individual stars are
implemented such that if either both or even just one of the components of a candidate binary fail the test, the binary candidate
is removed from consideration.

 \begin{figure}
 \vskip 0pt
 \hskip -5pt
 \includegraphics[height=7.0cm,width=8.5cm]{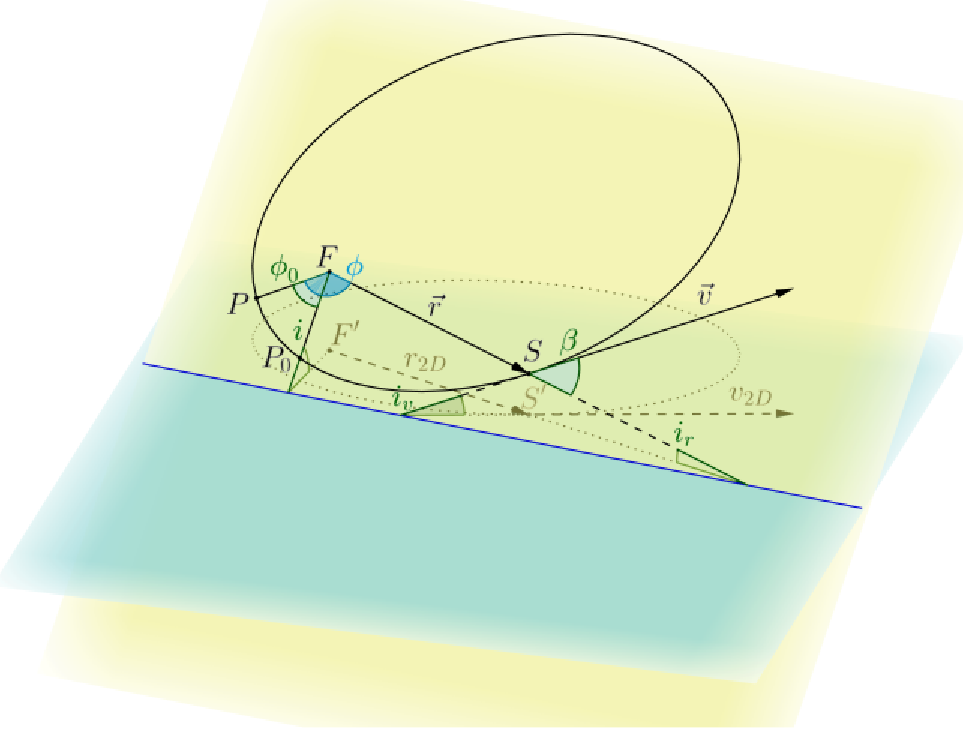}
 \caption{The figure illustrates the projection of the elliptical orbit described by the relative separation, $\vec{r}$, and relative velocity, $\vec{v}$,
   of both stars in the plane of their orbit, yellow, onto the plane of the sky, shown in blue, see text.}
 \end{figure}

Regarding the exclusion of flybys, having radial velocities for all stars, we exclude from consideration any binary candidate where the
difference between the radial velocity measurements of both components exceeds 4 km s$^{-1}$. Binaries with relative internal velocities
on the plane of the sky, $v_{2D}$, above 4 km s$^{-1}$ are likewise excluded. Given the pairwise relative velocity distribution of field
stars in the Solar Neighbourhood is a Gaussian with a $1\sigma$ value close to 60 km s$^{-1}$, and that the average interstellar
separation is close to 1pc, the expected number of flybys satisfying simultaneously $r_{2D}<0.06$pc and relative velocities both along
the line of sight and on the plane of the sky below 4 km s$^{-1}$, is negligible in our final sample.

{  Then, a signal-to-noise quality cut of 1.5 is applied to the resulting binary relative velocity values, with binaries where
the velocity signal-to-noise ratio in either RA or Dec is below this threshold are removed. This excludes cases where either of the two velocity
components are poorly measured, which occurs in 15\% of cases for each RA and Dec. As this \% is small, the chances of both RA and Dec components
having poorly measured relative velocities is quite small. This ensures no small relative velocity cut is being introduced, just a filter on cases
where either component of $v_{2D}$ is suspect. Thus, a final close to 34\% of cases were removed through this cut. After this cut, the average $v_{2D}$
signal-to-noise values are of $<v_{2D}/\sigma_{v}>=18.4$  when $r_{2D}<0.01$pc, and  $<v_{2D}/\sigma_{v}>=7.9$  when $r_{2D}>0.01$pc, much larger
than the 1.5 quality cut filter.} To further exclude the possibility that any non-Newtonian signal in the low acceleration $r_{2D}>0.01$pc
region might be the result of kinematic contaminants or noise, any binaries with $v_{2D}>1$km s$^{-1}$ are also removed, in this
region only e.g. Chae (2023a).

Resulting relative velocities in both RA and Dec. as a function of $r_{2D}$ are shown in Fig.(1). Binned mean values in these quantities are
shown by the circles with error bars, for RA and Dec. measurements, green and red, respectively. The red line gives a $v\propto r_{2D}^{-1/2}$
scaling fitted to the $r_{2D}<0.01$pc range, which as shown for a very similar clean sample in Hernandez (2023), is an accurate fit to Newtonian
expectations of Jiang \& Tremaine (2010). However, we see a regime change on crossing $r_{2D}=0.01$ pc, 2000 au, where the averaged binned velocity values
shift to another $v\propto r_{2D}^{-1/2}$ scaling, which appears slightly above. Reading this boost factor from the graph suggests an underlying model
where $G \to \gamma G$ on reaching the low acceleration $r_{2D}>0.01$pc regime, with a value of $\gamma=1.27$. This estimate is suggestive of the
$\gamma=1.43 \pm 0.06 $ reported by Chae (2023a), in good agreement also with MOND AQUAL expectations.

The estimate of $\gamma$ described above is crude for a number of reasons; the details of the fit are somewhat subjective to exactly which sets
of mean points are used for each fit, points which in turn depend on the details of the binning performed. Also, this potentially crucial gravitational
anomaly is being inferred from a discrete parameter of a complex velocity distribution, with the necessary lack of robustness and loss of information
intrinsic to any binning procedure.

For this reason in the following section we develop a formal probabilistic model to carefully include all
details of the inherent probability density functions (PDFs) at play: two for the relevant projection angles, one for a sampling of an orbital phase,
one for a distribution of semi-major axes and one for a sampling of an ellipticity distribution, all for a given observed set of $r_{2D}$, $v_{2D}$
values, and particular total binary masses, $M_{T}$. This will allow a formal testing of the $G \to \gamma G$ hypothesis and return best fit inferred
values of $\gamma$, both in the high acceleration $r_{2D}<0.01$ pc and in the low acceleration $r_{2D}>0.01$ pc regimes, paying attention to the details
of un-binned distributions of relative velocities, and no longer focusing on specific moments of these distributions. Full statistical, $\sigma_{st}$,
resolution, $\sigma_{re}$, and systematic, $\sigma_{sy}$, $1\sigma$ confidence intervals on inferred values of $\gamma$, will be developed and presented.

\section{Statistical framework}

 \begin{figure}
 \vskip 0pt
 \hskip -5pt
 \includegraphics[height=7.0cm,width=8.5cm]{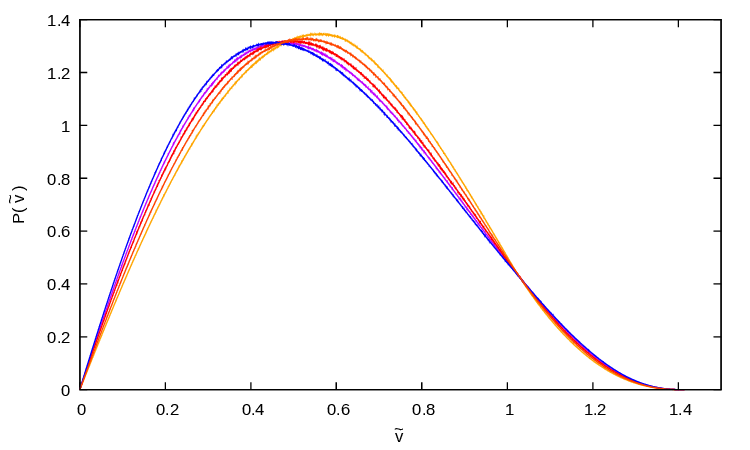}
 \caption{$P(\tilde{v})$ curves from 1.5$\times 10^{10}$ samples of equation (10) presented using 5000 $\tilde{v}$ bins for {  $\gamma=1$}. The
   various curves are for different assumed ellipticity distributions parameterised using $P(e)=(1+\alpha)e^{\alpha}$: $\alpha=0.6$, yellow, $\alpha=0.8$,
   orange, $\alpha=1.0$, thermal, red, $\alpha=1.2$, purple and $\alpha=1.4$, blue, respectively. {  For any arbitrary value of $\gamma$, these same
   curves can be trivially re-scaled using equation (10).}}
 \end{figure}

The model which we shall test is one where gravity is purely Newtonian, but where the actual value of the gravitational constant is allowed to
vary by a scale factor, such that $G \to \gamma G$. A full probabilistic model will be presented such that use of all information content of the data is
what determines the inferred value of $\gamma$ and its corresponding confidence interval, under a flat prior assumption which neither enhances nor
diminishes the probability of obtaining $\gamma=1$ either in the high acceleration $r_{2D}<0.01$ pc region, or in the low acceleration $r_{2D}>0.01$ pc
one. The orbits of the binary stars will hence be assumed to be Keplerian ellipses, and the sample will be assumed to be free of kinematic contaminants,
in accordance to the strict sample selection criteria described in the previous section. Systematics regarding a scenario where this last assumption
could be invalid, will be considered in the final section.

Each observed binary star, as described in the previous section, consists of two inferred masses, $m_{1j}$ and $m_{2j}$,
from which a total mass per binary of $M_{Tj}=m_{1j}+m_{2j}$ follows, a measured separation on the plane of the sky, $r_{2Dj}$, a relative velocity between
the two components on the plane of the sky, $v_{2Dj}$, and an error on this last quantity, $\sigma_{vj}$. With use of {\it Gaia} FLAME masses for most of
the stars included, and of magnitude-mass scalings calibrated using {\it Gaia} FLAME masses, uncertainties in the masses will be below 10$\%$, close to 5$\%$
on the average, Hernandez (2023), Chae (2023a). Given the upper distance of the sample of only 125pc, yielding average signal-to-noise values for the parallax
of our {\it Gaia} sample of close to 1000, {  with medians of 764.4 and 740.6 for the primaries and secondaries, respectively} (see Hernandez 2023), implies
that the errors on $r_{2D}$ and $M_{T}$ will be much smaller than those on $v_{2D}$. This is particularly relevant in the critical $r_{2D}>0.01$pc region where
final mean $<v_{2Dj}/\sigma_{vj}>=7.9$, 13$\%$ errors. Further, the dependence of velocity on only the square root of separation and the square root of mass
implies that adding in quadrature, errors on velocity based inferences will be dominated, by well over an order of magnitude, by the satellite reported
$\sigma_{vj}$ quantities. Hence, we shall consider these last errors fully and consistently in the statistical and probabilistic model constructed, and
ignore the errors on both $r_{2D}$ and $M_{T}$. A full validation of the entire scheme through the extensive use of synthetic samples will also be included.

From a Bayesian perspective, our first step is to calculate the probability that a given data point might arise from a particular model, i.e., from
a particular value of $\gamma$. We hence have to calculate the probability density distributions for both $r_{2D}$ and $v_{2D}$, given a value of $\gamma$.
These are obtained from the probability density functions determining the details of a binary orbit, and the projection of the relative velocity and separation
on the plane of the sky of both components, as follows:

\begin{figure*}
    \includegraphics[height=7.0cm,width=8.8cm]{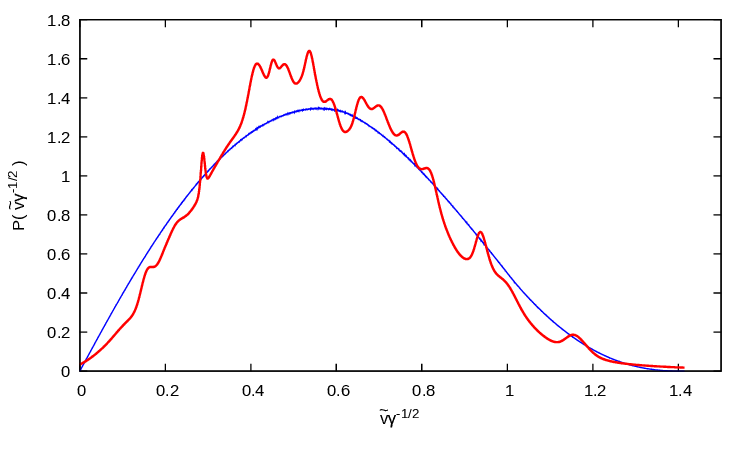}
     \hspace*{-5pt}
     \includegraphics[height=7.0cm,width=8.8cm]{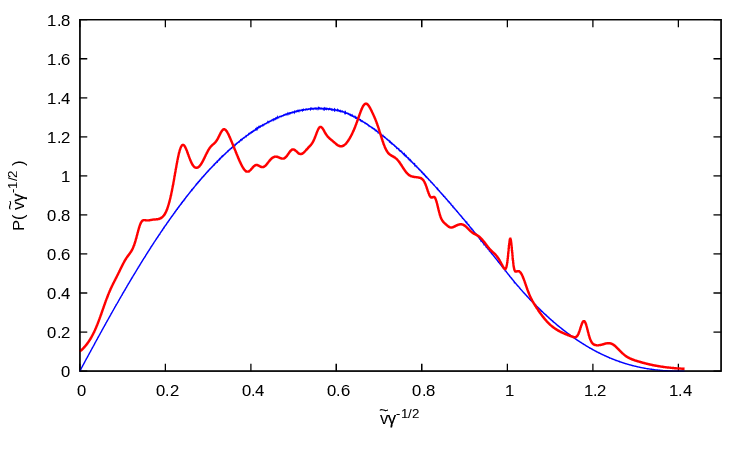}
    
     \caption{{ Left(a)}: the red curve shows the inferred $P_{I}(\tilde{v}\gamma^{-1/2})$ curve for the 466 data points in the $r_{2D}<0.01$pc range, for the
       parameters minimising the KS test distance, $\alpha=0.6$ and $\gamma=1.000$. The blue curve gives the theoretical $P(\tilde{v}\gamma^{-1/2})$
       curve of this optimal comparison, at $\alpha=0.6$.
       { Right(b)}: the red curve shows the inferred $P_{I}(\tilde{v}\gamma^{-1/2})$ curve for one particular of the 50 synthetic data samples
       produced, in the $r_{2D}<0.01$pc range, for the parameters minimising the KS test distance for the data, $\alpha=0.6$ and
       $\gamma=1.000$. All synthetic samples have identical sets of $r_{2Dj}$, $M_{Tj}$ and $\sigma_{vj}$ values as the observed sample and
       differ only in having $v_{2Dj}$ values obtained from a random sampling of equation (10). The blue curve is the same as in the
       left panel. All curves have a $\tilde{v}\gamma^{-1/2}$ resolution of $\sqrt{2}/5000$.}   
 \end{figure*}

For an inclination angle of the orbital plane of a binary system of $i$ to the line of sight, the inclination of the relative velocity
vector between both components, $\vec{v}$ will be $i_{v}$, where $\sin i_{v}=\sin i |\cos(\beta+\phi-\phi_{0})|$, and that of the relative
position of both components, $\vec{r}$, will be $i_{r}$, where $\sin i_{r}=\sin i |\cos(\phi-\phi_{0})| $, with $\beta$ the angle between $\vec{r}$
and $\vec{v}$, and $\phi_{0}$ the phase angle of the radius vector having the largest inclination. The geometric set-up of the problem is summarised
in Fig. (2), with the plane of the orbit shown in yellow, and the plane of the sky in blue. For a system with semi-major axis, $a$, total mass
$M_{T}=m_{1}+m_{2}$ and ellipticity $e$, the instantaneous relative velocity in 3D between the two components will be given by:

\begin{equation}
v^{2}=\frac{\gamma G M_{T}}{a} \left( \frac{2a}{r} - 1\right),
\end{equation}

\noindent where $r$ is given by:

\begin{equation}
r=\frac{a(1-e^{2})}{1+e\cos\phi},
\end{equation}

\noindent where $\phi$ is the true anomaly, the orbit phase angle measured from the pericentre. Using the last equation, equation (1) can be written as:

\begin{equation}
v^{2}=\frac{\gamma G M_{T}}{a} \left( \frac{1+e^{2}+2e\cos\phi}{1-e^{2}}  \right).
\end{equation}

Two auxiliary constant quantities which will be of use are the expressions for the magnitude of the angular momentum of the orbit, and the orbital period:

\begin{equation}
L=r v \sin \beta=[\gamma G M_{T} a(1-e^{2}) ]^{1/2},
\end{equation}

\begin{equation}
\tau=2 \pi a \left ( \frac{a}{\gamma G M_{T}}  \right)^{1/2}.
\end{equation}

We can now write the 2D projections of $\vec{v}$ and $\vec{r}$ using the inclination angle of the orbital plane, $i$, as:

\begin{equation}
r_{2D}=r\cos i_{r}=\frac{a(1-e^{2})}{1+e\cos \phi} \left[ 1-\sin^{2}i \cos^{2}(\phi-\phi_{0}) \right]^{1/2}
\end{equation}

\noindent and,

\begin{equation}
\begin{split}
  v_{2D}=v \cos i_{v}=\left( \frac{\gamma G M_{T}}{a}\right)^{1/2} \\
  \left[ \frac{(1+e^{2}+2e\cos\phi)(1-\sin^{2}i\cos^{2}(\beta+\phi-\phi_{0}))}{(1-e^{2})}\right]^{1/2}.
\end{split}
\end{equation}

For the angle between $\vec{v}$ and $\vec{r}$ we have:

\begin{equation}
\sin \beta=\frac{L}{rv}=\frac{1+e\cos \phi}{(1+e^{2}+2e\cos \phi)^{1/2}}.
\end{equation}

Since we are assuming $r_{2D}$ is an observed quantity with very little uncertainty for each nearby {\it Gaia} binary pair, we can eliminate the dependence
of $v_{2D}$ on $a$ by writing $v_{2D}$ in terms of $r_{2D}$:

\begin{equation}
\begin{split}
  v_{2D}= \left( \frac{\gamma G M_{T}}{r_{2D}}\right)^{1/2} \left[ 1-\sin^{2}i\cos^{2}(\phi-\phi_{0}) \right]^{1/4}\\
  \left[ \frac{1+e^{2}+2e\cos\phi-\sin^{2} i(e\sin\phi_{0}-\sin(\phi-\phi_{0}))^{2}}{1+e\cos\phi} \right]^{1/2}
\end{split}
\end{equation}

introducing $\tilde{v}=(G M_{T}/r_{2D})^{-1/2}v_{2D}$ we can write:

\begin{equation}
  \begin{split}
    \tilde{v}=\gamma^{1/2} \left[ 1-\sin^{2}i\cos^{2}(\phi-\phi_{0}) \right]^{1/4}\\
    \left[ \frac{1+e^{2}+2e\cos\phi-\sin^{2} i(e\sin\phi_{0}-\sin(\phi-\phi_{0}))^{2}} {1+e\cos\phi} \right]^{1/2}
  \end{split}
\end{equation}

A distribution function for $\tilde{v}$ can now be obtained from the above equation since the distribution functions for the angles involved are well known,
with isotropy implying $P(i)\propto \sin i$, $\phi_{0}$ being uniformly distributed between $0$ and $2 \pi$ and $e$ having a distribution function which we take
from the parametric form given in Hwang et al. (2022), $P(e)=(1+\alpha) e^{\alpha}$. The distribution function for the angle $\phi$ can now be obtained through
the time spent at each phase interval using $L=r^{2}d\phi/dt$ as follows:

\begin{equation}
\tau=\int_{0}^{\tau} dt =\int_{0}^{2\pi} \frac{r^{2} d\phi}{L} = \int_{0}^{2\pi} \frac{a^{2}(1-e^{2})^{2} d\phi}{L(1+e\cos\phi)^{2}}.
\end{equation}

\noindent Using equations(4) and (5) for $L$ and $\tau$ we get:

\begin{equation}
1=\int_{0}^{2\pi} \frac{(1-e^{2})^{3/2} d\phi}{2\pi(1+e\cos\phi)^{2}}= \int_{0}^{2\pi} P(\phi) d\phi  
\end{equation}

\noindent and therefore,

\begin{equation}
P(\phi)=\frac{(1-e^{2})^{3/2}}{2\pi(1+e \cos\phi)^{2}}
\end{equation}

Hence, distributions functions for two projection angles $i$ and $\phi_{0}$, for the true anomaly $\phi$, and for the ellipticity, $e$,
fully determine the probability density distribution of $\tilde{v}\gamma^{-1/2}$, for an assumed value of $\gamma$, from which an observed $r_{2D}$, and
observed $M_{T}$ and an assumed value of $\gamma$ will yield a PDF for $v_{2D}$. Or inversely, a set of observed $r_{2D}$ and $v_{2D}$, observed
$M_{T}$ values and an assumed value of $\gamma$ will yield an empirical $\tilde{v}\gamma^{-1/2}$ distribution. The PDF for $a$ will necessarily
be the one present in the data, through the elimination of $a$ in favour of $r_{2D}$ included in the use of eq.(2).

The PDF resulting from equation (10) is the master equation of the problem, and the one against which empirically inferred $P_{I}(\tilde{v})$ PDFs
will be compared. Unfortunately, this PDF does not seem to be analytical, in particular given the continuous dependence of $P(e)$ on $\alpha$, which we wish to leave
as a free parameter given recent evidence of continuous variations of the ellipticity distribution of {\it Gaia} wide binaries with $r_{2D}$ 
reported by Hwang (2022). Thus, we perform high quality numerical samplings of $P(\phi_{0})$, $P(i)$, $P(e)$ and $P(\phi)$ to obtain large samples of
$5.1\times 10^{10}$ values of $\tilde{v}$ from equation (10), which are then binned with a resolution of $\sqrt{2}/5000$ for values of $\alpha$ of
0.6, 0.8, 1.0 (corresponding to a thermal ellipticity distribution), 1.2 and 1.4, covering the range of values of $\alpha$ found by Hwang (2022) for
{\it Gaia} wide binaries in the $r_{2D}$ range covered by our sample. These curves are shown in Fig. (3) {  for $\gamma=1$}, with a colour code which
will be maintained throughout. {  For any arbitrary value of $\gamma$, these same curves can be trivially re-scaled using equation (10).} Notice the
two critical points in $\tilde{v}$ values where all $\alpha$ curves closely cross. This feature can be used in wide binary gravity tests when large samples
are involved, to eliminate systematic uncertainties due to unknown details in the ellipticity distribution and its possible $r_{2D}$ dependences.

Next we construct inferred $P_{I}(\tilde{v})$ functions as described below. Assuming Gaussian errors, the observation of a $v_{2Dj}$ value and
its accompanying $\sigma_{vj}$ confidence interval has to be viewed as a Gaussian PDF centred on $v_{2Dj}$ and having a standard deviation of $\sigma_{vj}$.
Hence, each observed binary will have associated to it a $\tilde{v}$ distribution given by:

\begin{equation}
P_{Ij}(\tilde{v})=\frac{1}{\sigma_{\tilde{v}j}\sqrt{2 \pi}}     e^{-(\tilde{v}-\tilde{v}_{j})^{2}/2\sigma^{2}_{\tilde{v}j}},
\end{equation}

\noindent where $\tilde{v}_{j}=(G M_{Tj}/r_{2Dj})^{-1/2} v_{2Dj}$ and $\sigma_{\tilde{vj}}=(M_{T}/r_{2D})^{-1/2}\sigma_{vj} $,
for a binary with an observed $r_{2Dj}$ value and inferred masses. A first empirical $P_{I}(\tilde{v})$ can now be constructed as:

\begin{figure*}
    \includegraphics[height=7.0cm,width=8.8cm]{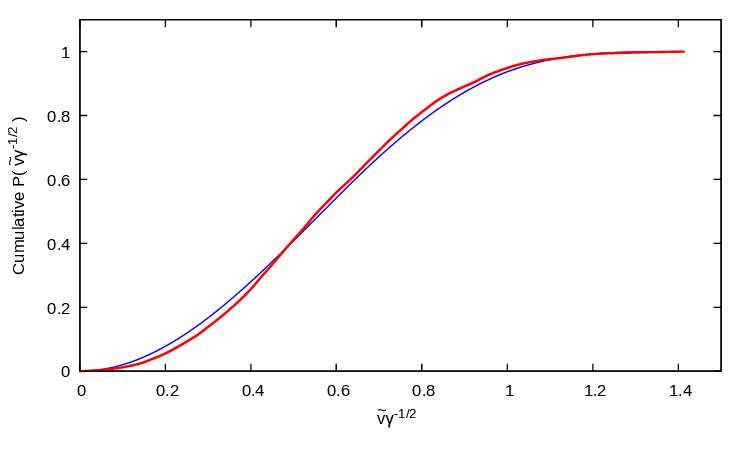}
     \hspace*{-5pt}
     \includegraphics[height=7.0cm,width=8.8cm]{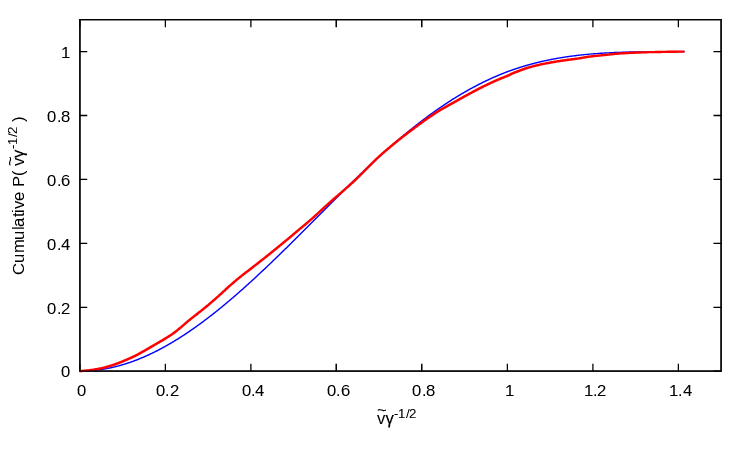}
    
     \caption{{Left(a)}: the red curve shows the cumulative distribution corresponding to the inferred $P_{I}(\tilde{v}\gamma^{-1/2})$ curve for the
       466 data points in the $r_{2D}<0.01$pc range, for the parameters minimising $D_{KS}$, $\alpha=0.6$ and $\gamma=1.000$.
       The blue curve gives the cumulative distribution corresponding to the theoretical $P(\tilde{v}\gamma^{-1/2})$ curve of this optimal comparison,
       at $\alpha=0.6$. {Right(b)}: the red curve shows the cumulative distribution corresponding to the inferred $P_{I}(\tilde{v}\gamma^{-1/2})$ curve
       for one particular of the 50 synthetic data samples produced, in the $r_{2D}<0.01$pc range, for the parameters minimising $D_{KS}$ for
       the data, $\alpha=0.6$ and $\gamma=1.000$. All synthetic samples have identical sets of $r_{2Dj}$, $M_{Tj}$ and $\sigma_{vj}$
       values as the observed sample and differ only in having $v_{2Dj}$ values obtained from a random sampling of equation (10). The blue curve
       is the same as in the left panel. All curves have a $\tilde{v}\gamma^{-1/2}$ resolution of $\sqrt{2}/5000$.}   
 \end{figure*}

\begin{equation}
P_{I}(\tilde{v})=\sum_{j=0}^{j=Ntot} P_{Ij}(\tilde{v}).
\end{equation}

\noindent The above procedure also has the advantage of eliminating any need for $\tilde{v}$ binning (e.g. Pittordis \& Sutherland 2023) when comparing
observed wide binary samples and theoretical $\tilde{v}$ distributions.

Since we are testing the hypothesis of Keplerian orbits and a fixed value of $\gamma$, the above function will be truncated at $\tilde{v}\gamma^{1/2}<0$ and
$\tilde{v}>\sqrt{2}\gamma^{1/2}$ and then normalised, as the Gaussian extensions of both low and high $\tilde{v}$ values will lead to unphysical tails at $\tilde{v}<0$
and $\tilde{v}>\sqrt{2}\gamma^{1/2}$. Finally, in the interest of sampling the range of $v_{2D}$ values consistent with the reported $\sigma_{vj}$ values, Gaussian
re-samplings of the original $v_{2Dj}$ values are performed to obtain alternative sets of $v_{2Dj}$ values at the same fixed $\sigma_{vj}$, $r_{2Dj}$ and $M_{Tj}$
values, with the final inferred $P_{I}(\tilde{v})$ curve being the average of a large sample of 500 such $v_{2Dj}$ re-samplings. This last step produces 
a mild smoothing of the $P_{I}(\tilde{v})$ curves, as the average signal-to-noise of the relative velocity values in our sample is of 15.7, necessary to
obtain well defined final goodness-of-fit curves, particularly for the small $N_{tot}$ $r_{2D}>0.01$ pc region where $<v_{2Dj}/\sigma_{vj}>=7.9$ and
total numbers are of only $N_{tot}=108$. Convergence was tested and negligible changes in all of the reported parameters resulted, for a range of 100-1000
such error re-samplings.

Once a $P_{I}(\tilde{v})$ has been constructed as described above for a given observed sample and an assumed value of $\gamma$, it can be compared to the
theoretical curves of $P(\tilde{v})$, for any desired value of $\alpha$. A sweep of values of $\gamma$ will then be performed for each relevant value
of $\alpha$, to obtain comparisons of the empirical $P_{I}(\tilde{v})$ curves to the master equation of the problem as a function of both $\gamma$
and $\alpha$. The comparison between an empirical $P_{I}(\tilde{v})$ and a theoretical  $P(\tilde{v})$ one will be carried out through a
Kolmogorov-Smirnov test, which yields a goodness-of-fit parameter as $(N_{tot})^{1/2}D_{KS}$, where $D_{KS}$ is the largest vertical difference at a fixed
$\tilde{v}$ value between the cumulative distributions being compared and $N_{tot}$ is the total number of observed wide binaries involved in any particular
comparison. This comparison was in practice performed using a $\sqrt{2}/5000$ $\tilde{v}\gamma^{-1/2}$ resolution.

Thus a maximum goodness-of-fit $\gamma_{BF}$ is obtained for every observed sample treated and every assumed value of $\alpha$. To obtain an internal
confidence interval on $\gamma_{BF}$ we proceed again through a Monte Carlo method, producing a synthetic wide binary sample having exactly the same
number of binaries, the exact same set of $r_{2Dj}$ values, same $M_{Tj}$ values, and same $\sigma_{vj}$ values, but having $v_{2Dj}$ values produced from a
random sampling of equation (10), at the maximum-goodness-of-fit $\gamma$ and $\alpha$ obtained for the observed sample. Hence, a statistical sample with
the same data and error structure of the best fit solution is produced, and treated exactly the same as the original data sample, yielding a corresponding
synthetic $\gamma_{SBF}$ best fit solution. This process is repeated 50 times to obtain a set of $\gamma_{SBFk}$ values, which are then found to be well
described by a Gaussian distribution, whose standard deviation becomes the statistical $1\sigma$ confidence interval of the original $\gamma_{BF}$ obtained
for the observed sample.

\begin{figure*}
    \includegraphics[height=7.0cm,width=8.8cm]{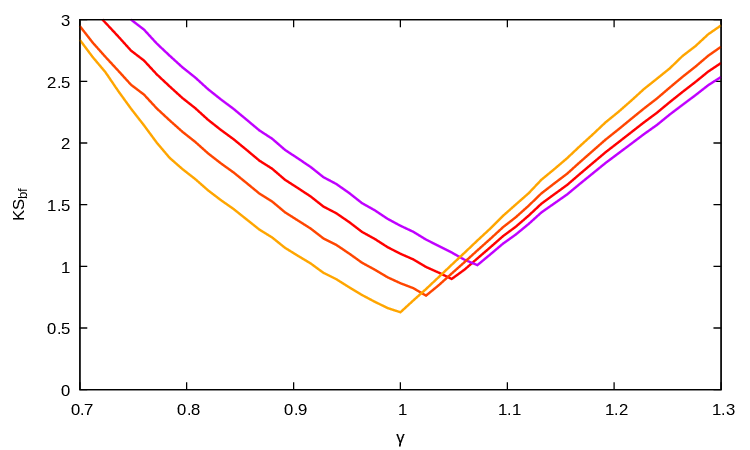}
     \hspace*{-5pt}
     \includegraphics[height=7.0cm,width=8.8cm]{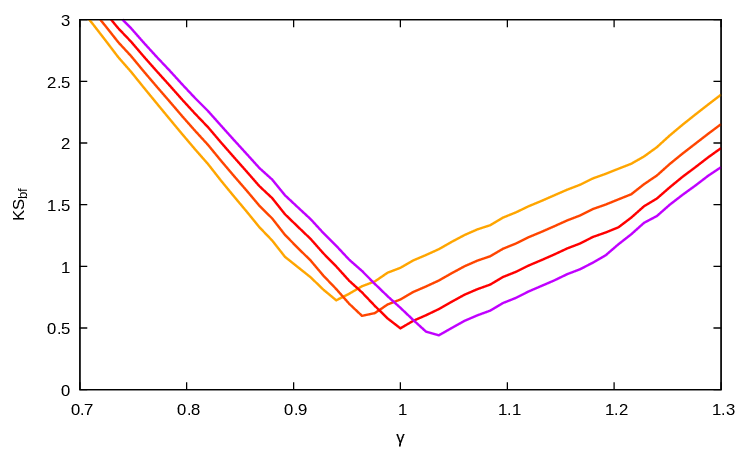}
    
     \caption{{Left(a)}: the figure shows KS test goodness-of-fit $\sqrt{N_{tot}} D_{KS}$ parameters for the 466 observed data point $P_{I}(\tilde{v}\gamma^{-1/2})$
       curves in the $r_{2D}<0.01$pc range, as a function of the assumed value of $\gamma$, for four different assumptions on the ellipticity
       distribution present, parameterised through $P(e)=(1+\alpha)e^{\alpha}$: $\alpha=0.6$, yellow, $\alpha=0.8$, orange, $\alpha=1.0$,
       thermal, red, $\alpha=1.2$ and purple, respectively. The data very clearly identify the Newtonian value of $\gamma=1$ as the optimal
       fit parameter. {Right(b)}: the figure shows KS test goodness-of-fit $\sqrt{N_{tot}} D_{KS}$ parameters for $P_{I}(\tilde{v}\gamma^{-1/2})$ curves for one
       particular synthetic sample produced using the best fit $\alpha=0.6$ and $\gamma=1.000$ parameters obtained for the observed data, in the
       $r_{2D}<0.01$pc range, as a function of the assumed value of $\gamma$, for four different assumptions on the ellipticity distribution present,
       parameterised through $P(e)=(1+\alpha)e^{\alpha}$: $\alpha=0.6$ -- yellow, $\alpha=0.8$ -- orange, $\alpha=1.0$ (thermal) -- red and $\alpha=1.2$ -- purple.
       The range of recovered best fit parameters from the full sample of 50 synthetic data realisations yields
       the internal statistical confidence interval for the $\gamma$ parameter recovered from the observed data of $\sigma_{st}= 0.054$, and of
       $\sigma_{sy}= 0.036$ due to systematic uncertainties in the ellipticity distribution, for the inferred $\gamma$. All curves have a
       $\gamma$ resolution of $0.012$.}   
 \end{figure*}

\section{Results}

\subsection{The $r_{2D}<0.01$ pc Sample}

We now describe the application of the statistical framework introduced in the previous section to the observed {\it Gaia} wide binary sample presented in
Section (2). We begin with the $r_{2D}<0.01$ high acceleration sub-sample, where equations (14) and (15) are used to produce an empirical
$P_{I}(\tilde{v}\gamma^{-1/2})$ PDF, for an assumed value of $\gamma$. This function is then compared through the KS test against the theoretical
$P(\tilde{v}\gamma^{-1/2})$ curves of Fig. (2), with the process repeated using a sweep of 50 evenly spaced values of $\gamma$ in the range $0.7<\gamma<1.3$.
An overall best fit value of $\gamma_{BF}=1.000$ is found for this sample at $\alpha=0.6$ for the theoretical curve.

The empirical $P_{I}(\tilde{v}\gamma^{-1/2})$ curve is shown by the red curve in the left panel of Fig.(4). The underlying discreteness of the sample
is still evident, despite having $N_{tot}=466$ binary pairs, with cases where the inferred confidence interval in $v_{2Dj}$, $\sigma_{vj}$, through reported
{\it Gaia} parameters is very small resulting in very narrow Gaussian distributions through equation (14) and hence the sharp peaks appearing in the curve
shown. The blue curve gives the best fit theoretical model at $\alpha=0.6$ and $\gamma_{BF}=1.000$.

The Kolmogorov-Smirnov comparison at the optimal parameters found is shown in the left panel of Fig.(5), which gives the cumulative distributions
corresponding to the curves shown in the left panel in Fig.(4), using matching colours, a good fit is obtained with a $\sqrt{N_{tot}}D_{KS}=0.628$.
The KS $\sqrt{N_{tot}}D_{KS}$ parameters of each of the 50 $\gamma$ sweeps for the values of $\alpha$ considered in the theoretical curves is presented in the
left panel of Fig.(6), where different colour curves correspond to different values of $\alpha$ in the theoretical $P(\tilde{v}\gamma^{-1/2})$ curves used in the
KS comparison against the inferred $P_{I}(\tilde{v}\gamma^{-1/2})$ coming from the data. We see all curves showing extremely well defined minima in all cases very close
to the Newtonian value of $\gamma=1$. We see also a small systematic offset appearing where assuming larger values of $\alpha$ leads to higher inferred
values of $\gamma$. As one assumes ellipticity distributions moving from sub-thermal to the $\alpha=1$ thermal case and beyond, higher ellipticities appear
leading to more elongated orbits. These more elongated orbits also imply stars spend longer periods of time at the large distance, slow moving phases
of their orbits, and hence the small shift in the $P(\tilde{v}\gamma^{-1/2})$ curves of Fig.(3) towards smaller values of $\tilde{v}$ as one goes to higher values of
$\alpha$. This effect in turn leads to larger values of inferred $\gamma$, as for a given observed set of $v_{2Dj}$ values, larger assumed values of
$\gamma$ will lead to smaller inferred values of $\tilde{v}\gamma^{-1/2}$.

There are strong theoretical expectations favouring the thermal ellipticity distributions of $\alpha=1$, e.g. Kroupa (2008), but also recent direct
observational determinations of this parameter precisely for {\it Gaia} wide binaries by Hwang et al. (2022). This last study, see their Fig. (7),
finds a value of $\alpha$ which varies from $\alpha=0.6$ to $\alpha=1.2$ for the ${r_{2D}}$ range covered by our $r_{2D}<0.01$ pc sample. Not wishing
to over-interpret this last result, which is the first published reference on the subject, we prefer not to modify the probabilistic model presented
to include an explicit variation of $\alpha$ with $r_{2D}$, but rather keep the ranges reported by Hwang et al. (2022) as an uncertainty range for this
parameter and add a systematic confidence interval due uncertainties in $\alpha$, to our inference procedure. This $\sigma_{sy}$ will be defined as half the
range in $\gamma_{BF}$ obtained over the range in $\alpha$ covered by the Hwang et al. (2022) results for the range in $r_{2D}$ covered by a particular data
set. For this first case, this systematic uncertainty will be of $\sigma_{sy}=0.036$. To this we must also add an uncertainty due to the resolution in the
implementation of the $\gamma$ sweep undertaken, of $\sigma_{re}=(0.6/50)/2=0.006$.

\begin{figure*}
    \includegraphics[height=7.0cm,width=8.8cm]{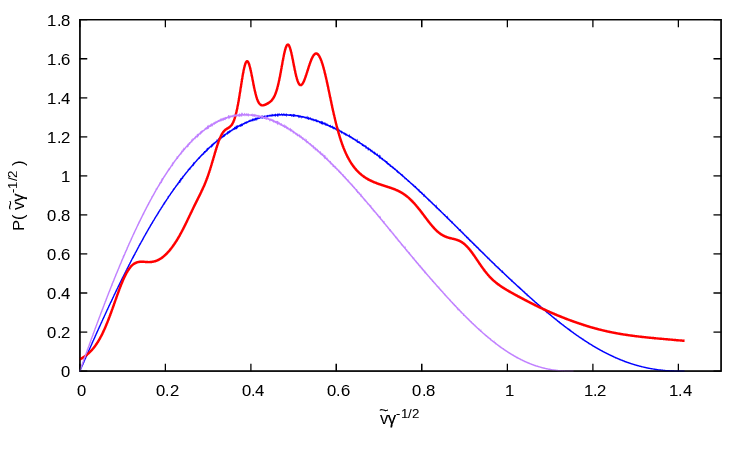}
     \hspace*{-5pt}
     \includegraphics[height=7.0cm,width=8.8cm]{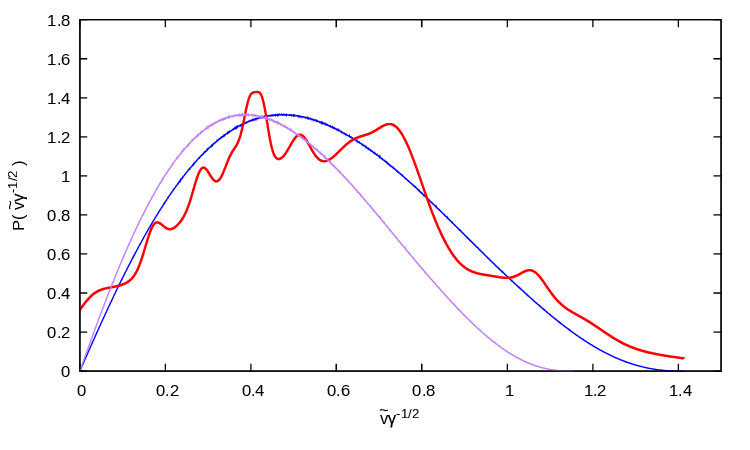}
     \caption{{ Left(a)}: the red curve shows the inferred $P_{I}(\tilde{v}\gamma^{-1/2})$ curve for the 108 data points in the $r_{2D}>0.01$pc range, for the
       parameters $D_{KS}$, $\alpha=1.2$ and $\gamma=1.512$. The blue curve gives the theoretical $P(\tilde{v}\gamma^{-1/2})$ curve of this optimal fit,
       at $\alpha=1.2$. For this same value of $\alpha$ the Newtonian model appears in purple, for comparison.
       { Right(b)}: the red curve shows the inferred $P_{I}(\tilde{v}\gamma^{-1/2})$ curve for one particular of the 50 synthetic data samples
       produced, in the $r_{2D}>0.01$pc range, for the parameters minimising $D_{KS}$ for the data, $\alpha=1.2$ and
       $\gamma=1.512$. All synthetic samples have identical sets of $r_{2Dj}$, $M_{Tj}$ and $\sigma_{vj}$ values as the observed sample and
       differ only in having $v_{2Dj}$ values obtained from a random sampling of equation (10). The blue and purple curves are the same as in the
       left panel. All curves have a $\tilde{v}\gamma^{-1/2}$ resolution of $\sqrt{2}/5000$.}   
 \end{figure*}

Lastly, to estimate the statistical uncertainty internal to the method, we turn to the Monte Carlo method as described in the previous section. Using as
input parameters the best fit $\gamma_{BF}=1.000$, $\alpha=0.6$ parameters found, and keeping the full $N_{tot}$, $r_{2Dj}$, $\sigma_{vj}$ and $M_{Tj}$ sets
of parameters of the observed sample, a set of 50 synthetic observations is produced where $v_{2D}$ is produced by sampling the best fit $P(\tilde{v}\gamma^{-1/2})$
curve at $\alpha=0.6$ and assuming $\gamma=1$. Each of these synthetic samples is then treated exactly as the original observational sample, to yield
a synthetic $\gamma_{BFS}$ value. These 50 different  $\gamma_{BFSk}$ values have a distribution which is well fitted by a Gaussian with a standard deviation
of 0.054, which hence becomes the internal statistical $1\sigma$ confidence interval of our method for the case considered, fully accounting for the
PDFs of the two projection angles of the problem, the sampling of the ellipticity distribution, the true anomaly distribution and the distribution
inherent to the velocity errors of the observed sample. This last sequence of obtaining synthetic observations is also repeated for the best fit $\gamma_{BF}$
values at the other $\alpha$ parameters considered, the resulting internal statistical errors are reported in Table (2).

As a final consistency check on the full method one can now check that the deviation between the centroids of the $\gamma_{BFSk}$
distributions are consistent with the input $\gamma_{BF}$ values to within the $1\sigma$ internal statistical confidence intervals found, which can be checked
to be the case in the last column of Table (2). The right panels of Figs.(4), (5) and (6) are analogous to the left panels, but show results for one particular
synthetic sample, at the overall best fit parameters of $\gamma=1.000$ and $\alpha=0.6$. This curves are seen to be qualitatively and quantitatively consistent
with the previous ones of the left panels, but do exhibit significant variations within this sample of 50 synthetic realisations, e.g. the actual values of
$\tilde{v}\gamma^{-1/2}$ at which particular sharp peaks occur shift from realisation to realisation, sometimes overlapping more, or less, while maintaining an overall
consistency, as seen in Fig.(5), where the real data curve of the left panel is actually a slightly better fit to the theoretical model than the particular
synthetic data set which was produced directly from sampling the model itself. In the right panel of Fig.(6) we see the particular synthetic sample presented
having an overall best fit at $\alpha=1.2$, with optimal KS values occurring at positions slightly displaced from those of the actual data sample shown in the
left panel of this figure. This small deviations are what give rise to the internal statistical and systematic errors inferred as described above.

As a final result for the inference of $\gamma$ for the $r_{2D}<0.01$pc we hence obtain: $\gamma=1.000 \pm \sigma_{st} \pm \sigma_{sy} \pm \sigma_{re}$
where $\sigma_{st}=0.054$, $\sigma_{sy}=0.036$ and $\sigma_{re}=0.006$ and therefore $\gamma=1.000 \pm 0.096$, a result fully consistent with Newtonian
expectations.

\subsection{The $r_{2D}>0.01$ pc Sample}

We now turn to the $r_{2D}>0.01$ pc sample, which will be treated in the same way as the previous one, with the only important difference being the
smaller number of observed binaries, which is now of only $N_{tot}=108$. Figs.(7), (8) and (9) are analogous to Figs.(4), (5) and (6), and show the
inferred $P_{I}(\tilde{v})$ for the real data, compared to the optimal model curve, the same comparison for the corresponding cumulative distributions,
and the $\sqrt{N_{tot}}D_{KS}$ values of all the $\gamma$ sweeps undertaken in the left panels. {  Fig. (7) differs slightly from Fig. (4) in that the Newtonian
model has been added in purple, separately from the best fit one in blue}. One extra step in this case, where smaller signal-to-noise
$v_{2D}$ values are present, is to take care that the overall data structure of the observed wide binary distribution is maintained. It can happen during
the velocity noise re-sampling phase that a small number of $v_{2D}$ values are shifted above the 1 km s$^{-1}$ upper limit imposed on the data in this
low acceleration $r_{2D}>0.01$ pc range as a safeguard against the inclusion of kinematic contaminants mentioned in section 2. Whenever this happens, the
data point in question is simply removed from consideration.

\begin{figure*}
    \includegraphics[height=7.0cm,width=8.8cm]{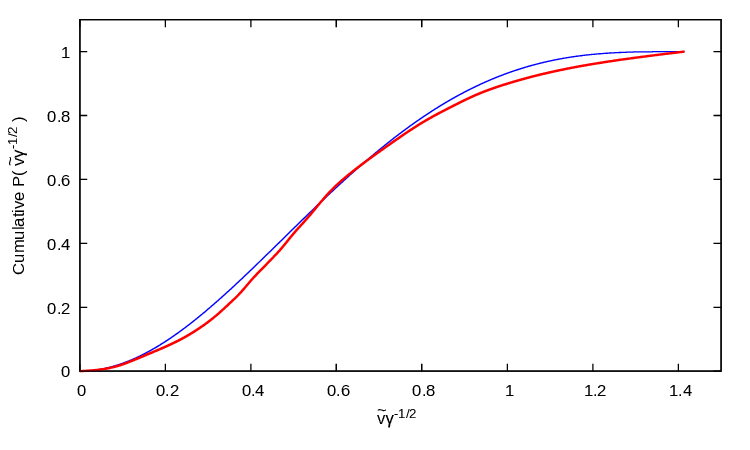}
     \hspace*{-5pt}
     \includegraphics[height=7.0cm,width=8.8cm]{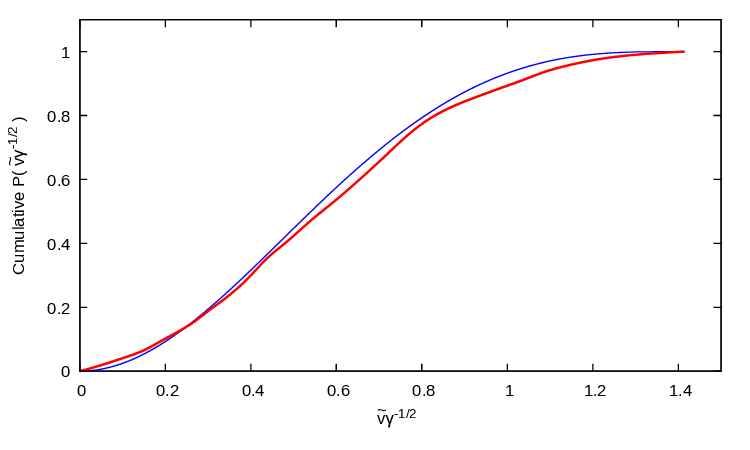}
    
      \caption{{Left(a)}: the red curve shows the cumulative distribution corresponding to the inferred $P_{I}(\tilde{v}\gamma^{-1/2})$ curve for the
       108 observed data points in the $r_{2D}>0.01$pc range, for the parameters minimising $D_{KS}$, $\alpha=1.2$ and $\gamma=1.512$.
       The blue curve gives the cumulative distribution corresponding to the theoretical $P(\tilde{v}\gamma^{-1/2})$ curve of this optimal comparison,
       at $\alpha=1.2$. {Right(b)}: the red curve shows the cumulative distribution corresponding to the inferred $P_{I}(\tilde{v}\gamma^{-1/2})$ curve
       for one particular of the 50 synthetic data samples produced, in the $r_{2D}>0.01$pc range, for the parameters minimising $D_{KS}$ for
       the data, $\alpha=1.2$ and $\gamma=1.512$. All synthetic samples have identical sets of $r_{2Dj}$, $M_{Tj}$ and $\sigma_{vj}$ values
       as the observed sample and differ only in having $v_{2D}$ values obtained from a random sampling of equation (10). The blue curve
       is the same as in the left panel. All curves have a $\tilde{v}\gamma^{-1/2}$ resolution of $\sqrt{2}/5000$.}   
 \end{figure*}

Given the results of Hwang et al. (2022), this time only three values of $\alpha$ were considered, which are the ones relevant for the $r_{2D}$ range our
second sample covers, $\alpha=1.0$, $\alpha=1.2$ and $\alpha=1.4$. Again, variations due to this range of ellipticity distributions will be considered a
systematic on the final results. As is natural due to the much reduced numbers involved, both the empirical $P_{I}(\tilde{v}\gamma^{-1/2})$ curve and the example
synthetic one, show stronger variations with respect to the underlying models than the previous sample. However, it is still the case that synthetic curves
produced from sampling the assumed underlying model are qualitatively and quantitatively analogous to the inferred curve produced from the real data,
as can be seen from the cumulative distributions of Fig.8.

In Fig.(9) we see  $\sqrt{N_{tot}}D_{KS}$ curves which are much noisier, but which still retain clearly defined optimal values, showing the same systematic
drift with assumed $\alpha$ as described for the previous sample. Indeed, as seen in the last column of Table (2), the consistency check on the method is
still positive, as the centroids of recovered $\gamma_{BFSk}$ samples are still well within the internal $1\sigma$ statistical confidence intervals of the
input parameters.

This time we obtain $\sigma_{st}=0.143$, $\sigma_{sy}=0.048$ and $\sigma_{re}=0.008$, for an overall best fit value of $\gamma_{BF}=1.5 \pm 0.2$. {  This
very substantial offset from Newtonian expectations is significantly larger than the uncertainties due to $\alpha$, as is evident when comparing the
small difference between $P(\tilde{v})$ curves for different values of $\alpha$ in Fig. (3) to Fig. (7), where the offset between the best fit model and the
Newtonian one is significantly larger. Thus, after fully accounting for all resolution, systematic and statistical confidence intervals,}
we see inferred values of $\gamma$ which are inconsistent with Newtonian expectations, at a $2.6 \sigma$ level. It is interesting that this result is
consistent with the value for this effective boost in $G$ recently reported by Chae (2023a) in the same $r_{2D}>0.01$pc range, using an independent approach
where hidden tertiaries are not removed from the sample, but included in the modelling of the observed internal relative velocities for a much larger
sample of close to 10,000 {\it Gaia} wide binaries. Chae (2023a) reports an inferred value of $\gamma=1.43 \pm 0.06$, in full consistency with results
presented here. As mentioned by Chae (2023a), these results are in close accordance with AQUAL expectations, a suggestion which is strongly reinforced
by the agreement of our results with those of Chae (2023a).

{  The use of full Monte Carlo simulations re-sampling the velocities while keeping the $r_{2D}$ values, masses, and crucially, the Gaia inferred $v_{2D}$
errors, explicitly probes the effects of the actual velocity errors present on the inference obtained. As seen in the second section of Table 2, inferred
values of $\gamma$ for the best fit parameters, particularly for the best fit $\alpha=1.2$ ($r_{2D}>0.01$ pc), are always well within $1 \sigma_{st}$ of
the input values, showing that the error structure of the data does not introduce any bias in the inference of $\gamma$. The only effect of the small
numbers and larger errors present in this region is a larger statistical confidence interval -- 2.4 times larger on average than what results for the
high acceleration region, where the sample is larger and relative errors smaller.

Notice that the overall velocity distributions, for both the high and low acceleration regions, remain consistent with the model expectations,
see Fig. 4, 5, 7 and 8, where no deficiency of low  $\tilde{v}$  cases are seen. The inferred values of $\gamma$ obtained are the result of the overall
distribution match, not of a low $\tilde{v}$  truncation. Having removed binaries with substantially higher noise level than the average has not biased
the velocity distributions away from the model expectations of elliptical orbits, but has indeed removed systems with poorly determined velocity
parameters. One of the causes of a poor proper motion determination is a poor single-stellar photometric, astrometric or photometric fit, which
in turn can be due to the presence of hidden tertiaries. Hence, all the data quality cuts introduced serve to limit the presence of any such contaminants.
Notice also the very close consistency of our results with the recent study of Chae (2023b), who also treats a small sample relatively
cleared of hidden tertiaries. Although the sample details of this last study differ from ours, the results are consistent.
}

Another consistency check can now be performed comparing results of the two samples presented as summarised in Table (2). From fundamental statistical
scalings, one should expect that the ratio between the $\sigma_{st}$ values obtained should scale close to inversely with the square root of the $N_{tot}$
numbers of both samples. The ratio of statistical $1\sigma$ confidence intervals for the two best fit parameters of the samples treated is of
$0.143/0.054=2.65$, while the square root of the ratio of wide binaries in these two samples is of $\sqrt{466/108}=2.08$, not far from the number above.

\begin{figure*}
    \includegraphics[height=7.0cm,width=8.8cm]{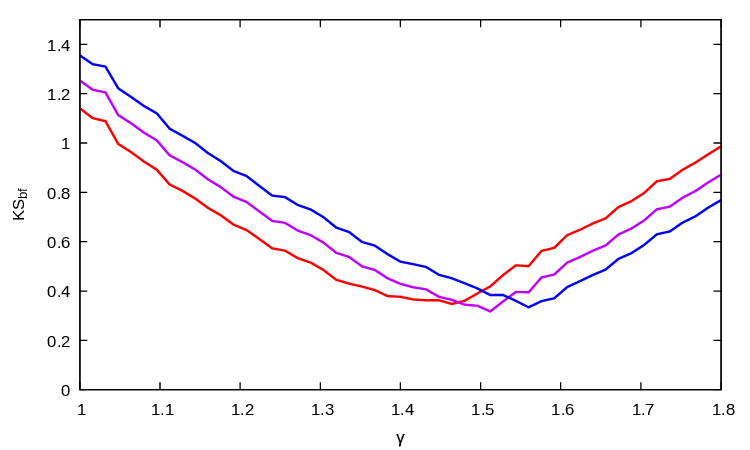}
     \hspace*{-5pt}
     \includegraphics[height=7.0cm,width=8.8cm]{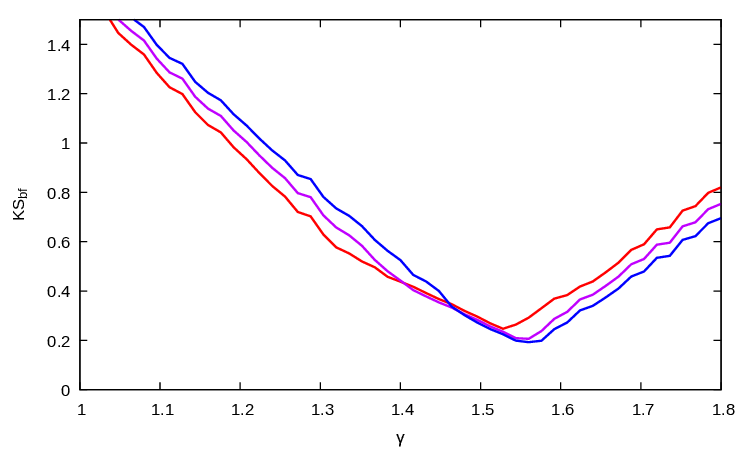}
    
      \caption{{Left(a)}: the figure shows KS test $\sqrt{N_{tot}} D_{KS}$ parameters for the 108 observed data points $P_{I}(\tilde{v}\gamma^{-1/2})$ curves
       in the $r_{2D}>0.01$pc range, as a function of the assumed value of $\gamma$, for three different assumptions on the ellipticity
       distribution present, parameterised through $P(e)=(1+\alpha)e^{\alpha}$: $\alpha=1.0$, thermal, red, $\alpha=1.2$, purple and
       $\alpha=1.4$, blue, respectively. The data very clearly identify the Newtonian values of $\gamma>1.3$ as optimal
       fit parameters. {Right(b)}: the figure shows KS test $\sqrt{N_{tot}} D_{KS}$ parameters for $P_{I}(\tilde{v}\gamma^{-1/2})$ curves for one particular synthetic
       sample produced using the best fit $\alpha=1.2$ and $\gamma=1.512$ parameters obtained for the observed data, in the $r_{2D}>0.01$pc
       range, as a function of the assumed value of $\gamma$, for three different assumptions on the ellipticity distribution present,
       $\alpha=1.0$ (thermal) -- red, $\alpha=1.2$ -- purple and $\alpha=1.4$ -- blue. The range of recovered best fit parameters
       from the full sample of 50 synthetic data realisations yields the internal statistical confidence interval for the parameters
       recovered from the observed data, of $\sigma_{st}= 0.143$, and of $\sigma_{sy}= 0.048$ due to systematic uncertainties in the ellipticity
       distribution, for the inferred $\gamma$. All curves have a $\gamma$ resolution of $0.016$.}   
    
 \end{figure*}

We end this section with Fig.(10) which summarises the results for the two samples considered, showing $\gamma_{BF}$ values and the sum of their internal
statistic and resolution $1\sigma$ confidence intervals, as a function of the assumed values of $\alpha$, for both samples discussed.

\section{Discussion}

Given the validation of the method presented through thorough checks using synthetic data samples and the perfect agreement with well established
Newtonian gravity $\gamma=1$ in the high acceleration $r_{2D}<0.01$pc regime, the results obtained for the $r_{2D}>0.01$pc region become important.
These show a clear $\gamma=1.5 \pm 0.2$ non-Newtonian behaviour of gravity in the low acceleration $r_{2D}>0.01$ pc region, corresponding to
an $a \approx 4 a_{0}$ threshold for the mean binary masses of the sample, Hernandez (2023). Though not strongly conclusive at a $2.6 \sigma$ level,
our results become compelling given the very close qualitative and quantitative agreement with the independent assesment of Chae (2023a) and Chae (2023b),
and are highly suggestive of MOND, given the AQUAL expectation for $\gamma \simeq 1.4$ for the wide binaries treated.

Despite the stringent clearing of all kinematic contaminants from the sample used, a procedure which has been validated in Hernandez et al. (2022)
and Hernandez (2023), and the recent direct observational results of Hwang et al. (2022), one can explore to what level our low acceleration
$\gamma >1$ result might arise from a failure of said cleaning methods and/or from the validity of the ellipticity distributions used. To this end
we now produce a synthetic $\gamma=1$ sample using the data structure of our 108 binary $r_{2D}>0.01$ pc sample, using an ellipticity
distribution parameter of $\alpha=0.6$, and run the inference method assuming an $\alpha=1.4$ model. Thus, we explore the maximum systematic offset
that reasonable uncertainties in the $\alpha$ parameter might induce. Also, allowing for some presently unknown mistake in the {\it Gaia} catalogue
which might result in the current reported confidence intervals being underestimated, we take $v_{2Dj}$ errors four times larger than what results
from standard error propagation analysis, $\sigma_{vj}=4\times \sigma_{vj}$. This will boost the average $v_{2D}$ values of the sample, as small
$v_{2Dj}$ cases will mostly end up at higher $|v_{2D}|$ values, and again bias the reconstruction procedure towards a higher $\gamma_{BF}$ region.

The results of this experiment are shown in Fig.(11) where the three panels are analogous to the left panels of Figs.(4), (5) and (6). An inferred
value of $\gamma_{BF}=1.23 \pm 0.148$ resulted. Not only is this still inconsistent with results from the real data sample in the low acceleration
region of $\gamma =1.5 \pm 0.2$, but even though a higher value of $\gamma$ resulted, the qualitative and quantitative structure of the inferred
$P_{I}(\tilde{v})$ is inconsistent with what is seen for the data. In the left panel of Fig. (11) we see the synthetic curve deviates more strongly from
the best-fit model than those in the preceding section, the much enhanced noise level shifting the curve towards a flatter distribution with a much
less well defined peak. This is confirmed in the next two panels, where the comparison is markedly poorer than when using the real data sample, with
$\sqrt{N_{tot}} D_{KS}$ values which grow by a factor of more than 3. Therefore, we see that the detailed distribution comparisons performed allow to
distinguish and flag samples where the input assumptions deviate from the model. This final test is much more discordant in the details than either the
data sample, or any of the synthetic samples produced in the previous section, and even so, pushing this $\gamma=1$ model to the limit only allows to
reach $\gamma=1.23$.

\begin{figure*}
    \includegraphics[height=7.0cm,width=8.8cm]{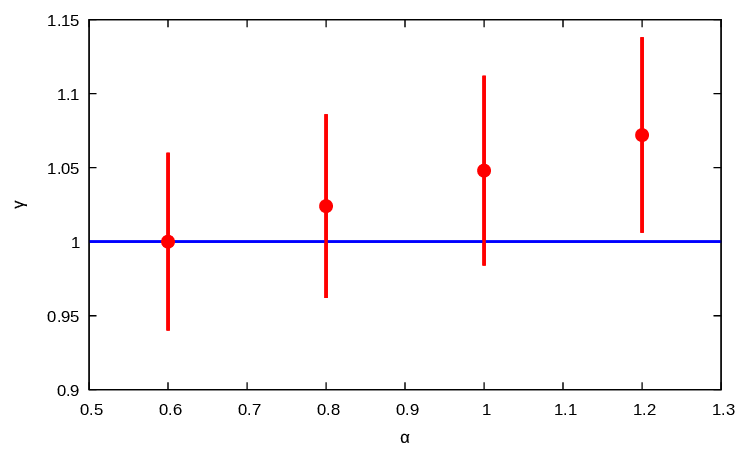}
     \hspace*{-5pt}
     \includegraphics[height=7.0cm,width=8.8cm]{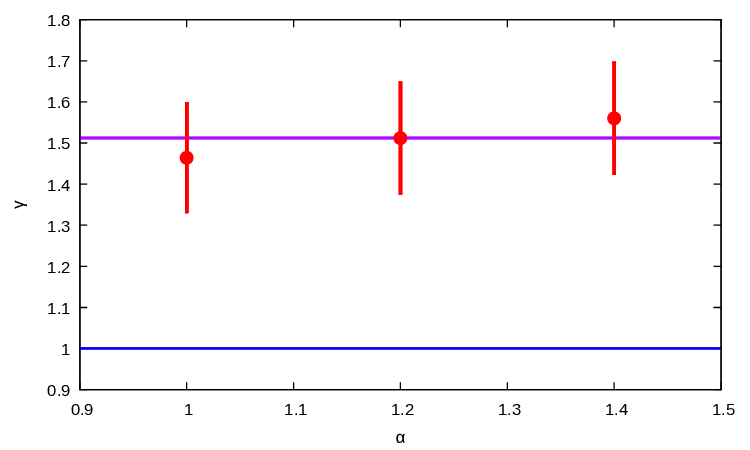}
    
     \caption{{Left(a)}:the figure shows best fit values of $\gamma$ for the $r_{2D}<0.01$ pc observed data sample as a function of the assumed value
       of $\alpha$. The vertical bars give $1\sigma$ confidence intervals through the addition of the standard deviation of the sample of
       recovered values of $\gamma$ from 50 Monte Carlo synthetic data samples constructed at the best fit parameters found for the data, and
       one half of the $\gamma$ resolution of the implementation. The horizontal blue line gives the Newtonian value of $\gamma=1$. Full consistency of
       our inferences with this value is evident for this first data sample.
       {Right(b)}: the figure is analogous to the left panel, but for the $r_{2D}>0.01$ pc data sample. Our inferences are inconsistent with the
       the horizontal blue line showing the Newtonian value of $\gamma=1$ at a 2.6$\sigma$ level, and much more suggestive of
       AQUAL expectations of close to $\gamma=1.4$.
       }   
 \end{figure*}

Although not explicitly included in this test, the presence of hidden tertiaries acts in a very similar way to noise, since the extra velocity
component of an inner binary sometimes adds and sometimes subtracts from the wide binary relative velocity, depending on the orientation of
the inner binary orbit with respect to the wide binary one. Hidden tertiaries hence increase mean $v_{2D}$ values while modifying the overall
velocity distributions, much as noise does. Indeed, Pittordis \& Sutherland (2023) explicitly note the degeneracy between an assumed hidden
tertiary fraction and an assumed flyby fraction, where flybys are modelled through a random sampling of asymptotic hyperbolic relative velocities,
again, much like noise, when comparing results against observed $\tilde{v}$ distributions.

{  A last caveat to mention is the possible presence of a fraction of hidden tertiaries in our sample, which can not be conclusively
rejected at this point, despite the very careful exclusion strategies implemented. Any such presence would bias results towards larger values of
$\gamma$, growing in relevance towards larger binary separations. We do stress that the consistency of the full $\tilde{v}$ distributions found with
theoretical expectations for pure elliptical binaries, see Figs. 4, 5, 6 and 7, argues against any significant remaining hidden tertiary fraction.
Similarly, the recent results of Chae (2023b), where an independent hidden tertiary cleansing scheme was applied to a similar {\it Gaia} wide binary
sample, yielding results consistent with no remaining hidden tertiaries in the high acceleration $r_{2D}<0.01$pc region, and consistent with
our results of $\gamma=1.5 \pm 0.2$ for $r_{2D}>0.01$pc, argue in the same direction. Notice the lack of any evidence for a change in the hidden
tertiary fraction with binary separation in the regime of relevance, e.g. Tokovinin et al. (2002) and Tokovinin, Hartung \& Hayward (2010).
It is fortunate that the resent results of Manchanda et al. (2023) show that any such hidden tertiary fraction can be found or discarded with current
available observational follow-up techniques,  by either future astrometric accelerations in the 10-year Gaia data, and/or speckle or coronagraphic
imaging on 8m telescopes, making a definitive settling of this point possible in the near future.}

The final test described in this section strongly suggest we are seeing a modified gravity phenomenology. Whilst the assumption of Newtonian or GR gravity
refers to very precise theories, modified gravity models, particularly covariant extensions to GR, appear in a great variety of forms and flavours.
To cite but a few, Milgrom (1983), Bekenstein (2004), Moffat \& Toth (2008), Zhao \& Famaey (2010), Capozziello \& De Laurentis (2011), Verlinde (2016),
Barrientos \& Mendoza (2018), Hernandez et al. (2019b) or Skordis \& Z\l o\'{s}nik (2021), all following distinctly different theoretical approaches.

One interesting conclusion of our results is that the transition between the Newtonian and the modified gravity regimes appears to
be fairly abrupt. Even for a discontinuous transition in gravitational regime with acceleration, this transition will appear smoother in the data analysed,
due to the unavoidable presence of wide binary stars with orbits that cross this transition. Yet, in the data analysed no intermediary transition regime
is evident. This could well be due to the poor sampling given the small numbers of wide binaries remaining in the very clean samples used, or to an actual
abrupt transition, e.g. as suggested by schemes where the change in gravity at low accelerations stems from quantum effects, e.g. Capozziello \&
De Laurentis (2011). In terms of particular models, it is clear that our results are closely consistent with MOND AQUAL $\gamma=1.4$ predictions e.g.
Chae (2023a).

Two points are identified as crucial towards increasing the precision of our inferences: an increase in the numbers of wide binaries used,
and a reduction of the systematic uncertainties through an improved empirical and theoretical understanding of the ellipticity distribution of
the binaries used, and its possible variations with $r_{2D}$. It is thus clear that future {\it Gaia} data releases will help significantly
towards a definitive answer from wide binary gravity tests, which presently lean towards modified gravity scenarios where there is no need to
invoke hypothetical dark matter components to understand galactic dynamics.

\begin{figure*}
    \includegraphics[height=7.0cm,width=5.9cm]{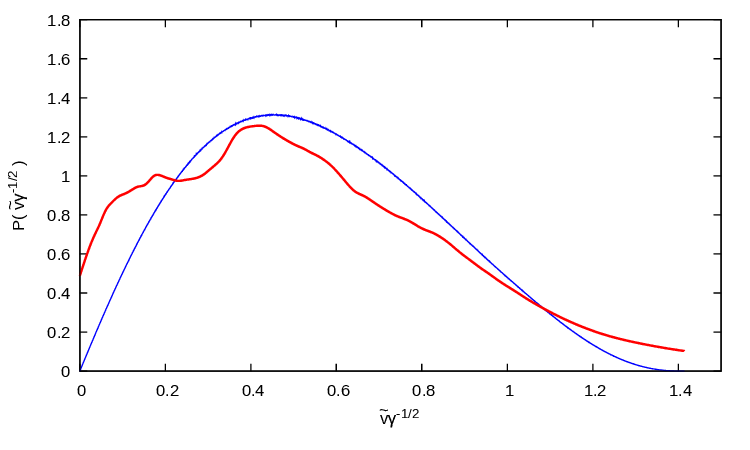}
     \hspace*{-5pt}
     \includegraphics[height=7.0cm,width=5.9cm]{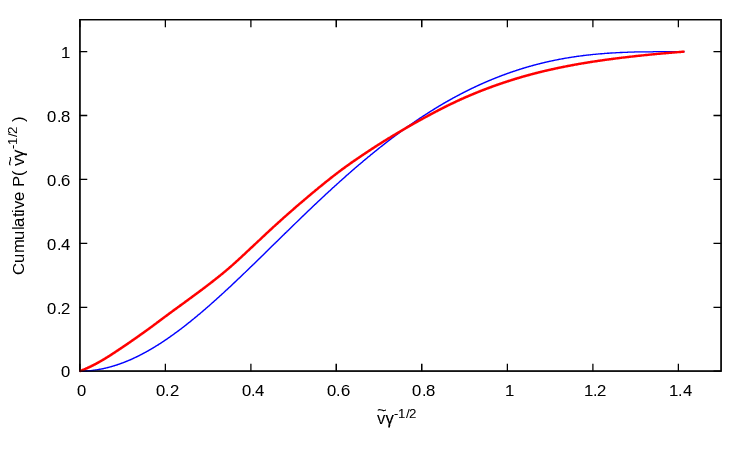}
      \hspace*{-5pt}
     \includegraphics[height=7.0cm,width=5.9cm]{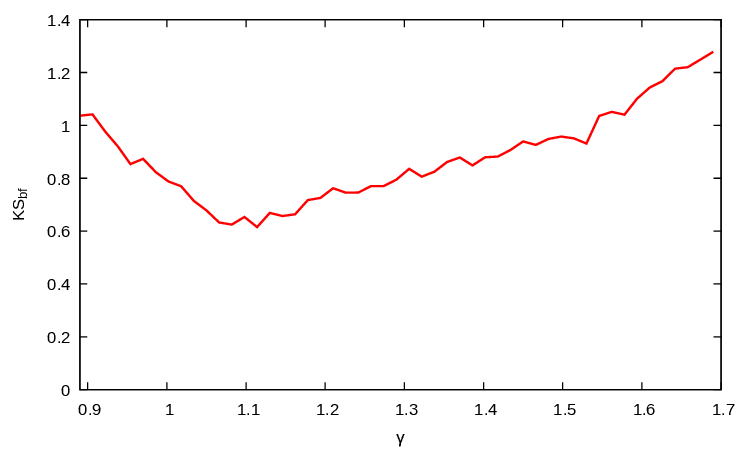}
    
     \caption{In an effort to reproduce a high inferred value of $\gamma_{BF}$ for the $r_{2D}>0.01$pc observed data sample within Newtonian assumptions, a set
       of synthetic Monte Carlo samples were probed with an input value of $\gamma=1.0$, an input value of $\alpha=0.6$, and compared against a
       $P(\tilde{v}\gamma^{-1/2})$ curve for $\alpha=1.4$, to maximise the systematic offset towards higher inferred values of $\gamma$. Also, the level of noise assumed
       was multiplied by a factor of 4 with respect to that reported in the {\it Gaia} data, to obtain larger mean $v_{2d}$ values and hence higher inferred
       values of $\gamma_{BF}$. Despite pushing the input $\gamma=1.0$ model to the limit, $\gamma_{BF}$ only reached values of 1.23. The three panels are
       analogous to those of Figs. (4), (5) and (6). The functional miss-match between the model constructed and the comparison $P(\tilde{v})$ model is evident,
       the best fit $P_{I}(\tilde{v}\gamma^{-1/2})$ curves are a very poor fit to the comparison model, with KS parameters being larger by a factor of 3 with respect to
       Fig. (9). Even though the inferred $\gamma_{BF}$ increases, the quality of the fit allows one to dismiss such an option where noise, systematic errors on
       the assumed ellipticity distribution and kinematic contaminants dominate.}   
 \end{figure*}

\section*{acknowledgements}

The authors acknowledge the input of the referee, Will Sutherland, as important towards having reached a more
balanced and complete final version. 
This work has made use of data from the European Space Agency (ESA) mission {\it Gaia} ({https://www.
cosmos.esa.int/gaia}), processed by the {\it Gaia} Data Processing and Analysis Consortium (DPAC, {https://www.
cosmos.esa.int/web/gaia/dpac/consortium}). Funding for the DPAC has been provided by national institutions, in
particular the institutions participating in the {\it Gaia} Multilateral Agreement. {\it Gaia} data retrieval
and initial processing (up to figure 1) was performed using software developed jointly with Stephen Cookson.
Xavier Hernandez and Alex Aguayo acknowledge financial assistance from CONAHCYT and PAPIIT IN102624. L. Nasser
gratefully acknowledges the support from the NSF award PHY - 2110425.

\begin{table*}
\begin{flushleft}
  \caption{Parameters for the two {\it Gaia} wide binary $r_{2D}$ cuts described.}
  \begin{tabular}{ | l | c c c c c c c c c | }
  \hline
  \hline   
 Projected  & Number      & Distance & $<D/pc>$ & $<S/N>_{v2D} $ &  $<S/N>_{\varpi}$  & $<RUWE>$ & $<B_{P}>$ & $<M_{T}/M_{\odot}>$  \\
 separation & of binaries & limit    &          &                   &                  &          &          &                    \\
 range      & used        & in pc    &          &                   &                  &          &          &                    \\   
   \hline
 $r_{2D}<0.01$pc & $466$   &  $125$    &  $86.9$  & $18.4$            & $901$            & $1.01$   & $0.12$    & $1.52$           \\
 $r_{2D}>0.01$pc & $108$   &  $125$    &  $87.1$  & $7.90$            & $892$            & $1.00$   & $0.12$    & $1.54$           \\

 \hline 
 
\end{tabular} 

  For the $r_{2D}<0.01$pc and $r_{2D}>0.01$pc samples described in the text the first three entries of the table give the number of binaries
  contained, the distance limit, and the mean distance of the sample. The next two entries give the mean signal-to-noise values for the
  2D relative internal velocities of the binaries and for the parallaxes of all stars used. The last three entries show average values for
  the {\it Gaia RUWE} parameter, the {\it Gaia} binary probability parameter for each individual star, and the mean binary masses.   
\end{flushleft}
\end{table*}

\begin{table*}
\begin{flushleft}
  \caption{Parameters of the two inference fits on the two wide binary selection cuts described.}
  \begin{tabular}{ | l | c c c c c c c c | }
  \hline
  \hline   
 Ellipticity   & $\gamma_{BF}$ & $(N_{tot})^{1/2}D_{KS}$ & $\sigma_{st}$ & $\sigma_{sy}$ & $\sigma_{re}$ & $\gamma_{BFsy}$ & $|\gamma_{BFsy}-\gamma_{BF}|/(\sigma_{st}+\sigma_{re})$ \\
 distribution  &              &                      &                 &              &               &                 &             \\
 parameter     &              &                      &                 &              &               &                 &             \\   
 \hline
 & & & & $r_{2D}<0.01$pc & & & \\
 \hline
 $\alpha=0.6 $    &  $1.000$    &  $0.628$        & $0.054$            & $0.036$      & $0.006$       & $0.956$          & $0.73$                     \\
 $\alpha=0.8 $    &  $1.024$    &  $0.763$        & $0.056$            & $0.036$      & $0.006$       & $1.000$          & $0.39$                     \\
 $\alpha=1.0 $    &  $1.048$    &  $0.898$        & $0.058$            & $0.036$      & $0.006$       & $1.044$          & $0.06$                     \\
 $\alpha=1.2 $    &  $1.072$    &  $1.092$        & $0.060$            & $0.036$      & $0.006$       & $1.092$          & $0.30$                     \\
 \hline
 & & & & $r_{2D}>0.01$pc & & & \\
 \hline
 $\alpha=1.0 $    &  $1.464$    &  $0.348$        & $0.120$            & $0.048$      & $0.008$       & $1.491$       & $0.21$                     \\
 $\alpha=1.2 $    &  $1.512$    &  $0.317$        & $0.143$            & $0.048$      & $0.008$       & $1.524$       & $0.08$                     \\
 $\alpha=1.4 $    &  $1.560$    &  $0.334$        & $0.147$            & $0.048$      & $0.008$       & $1.651$       & $0.59$                     \\
 \hline

\end{tabular} 

  For the $r_{2D}<0.01$pc and $r_{2D}>0.01$pc observational samples described in the text, the first two entries give the best fit inferred values of the
  factor multiplying $G$ to give the effective value of $\gamma G$ found, and the Kolmogorov-Smirnov function of merit for the optimal parameter found at
  each assumed value of $\alpha$, the parameter describing the assumed ellipticity distribution function. Also as a function of assumed $\alpha$, the next
  entry gives the results of the Monte Carlo synthetic samples produced at the optimal inferred parameters for the observed binaries, from which a
  distribution of synthetic inferred values of $\gamma$ appears having a distribution well described by a Gaussian, centred at $\gamma_{BFsy}$ and having
  a $1\sigma$ confidence interval of $\sigma_{st}$. $\sigma_{sy}$ gives the systematic errors in our $\gamma$ inferences due to the uncertainty in the
  details of the relevant ellipticity distributions. $\sigma_{re}$ gives one half of the $\gamma$ resolution of the implementation. The final entry shows
  the offset between the centroid of the distribution of recovered values of $\gamma$ from the synthetic samples constructed at the optimal parameters
  recovered from the data, in units of the total internal confidence interval. That this final numbers are consistently below 1 validates the full
  method presented.

\end{flushleft}
\end{table*}

\section*{DATA AVAILABILITY}
All data used in this work will be shared on reasonable
request to the author.

\end{document}